\documentclass[preprint,english,12pt]{elsarticle}
\usepackage{layout}
\usepackage[latin1]{inputenc}
\usepackage{babel}
\usepackage[T1]{fontenc}
\usepackage{amsmath, latexsym, soul, color, type1cm,dsfont, float, fancyhdr}
\usepackage[official]{eurosym}
\usepackage{graphicx}
\usepackage{abstract}
\usepackage{amssymb}

\journal{Annals of Physics}

\begin{document}


\title{A compact 341 model at TeV scale}
\author[label1]{A. G. Dias}
\ead{alex.dias@ufabc.edu.br}
\author[label2]{P. R. D. Pinheiro}
\ead{prdpinheiro@fisica.ufpb.br}
\author[label2]{C. A. de S. Pires}
\ead{cpires@fisica.ufpb.br}
\author[label2]{P. S. Rodrigues da Silva\corref{cor1}}
\ead{psilva@fisica.ufpb.br}
\cortext[cor1]{Corresponding author}
\address[label1]{Centro de Ci\^encias Naturais e Humanas, Universidade Federal
do ABC, R. Santa Ad\'elia 166, 09210-170, Santo Andr\'e - SP, Brazil.}
\address[label2]{Departamento de F\'{\i}sica, Universidade Federal da Para\'{\i}ba, 
Caixa Postal 5008, 58051-970, Jo\~ao Pessoa - PB, Brazil.}
\date{\today}

\maketitle
\begin{abstract}
We build a gauge model based on the $SU(3)_c\otimes SU(4)_L\otimes
U(1)_X$ symmetry where the scalar spectrum needed to generate gauge boson
and fermion masses has a smaller scalar content than usually assumed in literature.
We compute the running of its abelian gauge coupling and show that a Landau pole shows up at the TeV scale, a fact that we use to consistently implement those fermion masses that are not generated by Yukawa interactions, including neutrino masses. This is appropriately achieved by non renormalizable effective operators, suppressed by the Landau pole scale.  Also, $SU(3)_c\otimes SU(3)_L\otimes U(1)_N$ models embedded in this gauge structure are bound to be strongly coupled at this same energy scale, contrary to what is generally believed, and neutrino mass generation is rather explained through the same effective operators used in the larger gauge group. Besides, their nice features, as the existence of cold dark matter candidates and the ability to reproduce the observed standard model Higgs-like phenomenology, are automatically inherited by our model. Finally, our results imply that this model is constrained to be observed or discarded soon, since it must be realized at the currently probed energy scale in 
LHC.

\end{abstract}
    
    
\section{Introduction}
\indent The recent discovery at the Large Hadron Collider (LHC) of a spin zero resonance~\cite{LHC}, most certainly the standard model (SM) Higgs boson, have already given some clues of what kind of new physics we need to proceed beyond.
Besides, we know that the SM is not the last
answer to the questions brought by the huge endeavor carried over the last three decades
of researching in particle physics. Some of the new challenges involve neutrino's mass
and mixing, the dark matter problem, the matter-antimatter asymmetry, among others.
This motivates us to search for new broader models and theories for attacking
these puzzles in the most simple and easiest way.
One such possibility is the enlargement of the gauge group symmetry, implying in new 
interactions and particle spectrum. Here we focus on one of these gauge extensions of the SM,
which can be realized at some TeV scale, the $SU(3)_c\otimes SU(4)_L\otimes U(1)_X$~\cite{Vol,341vicente}
gauge group, which we call 341 for short.

The 341 model is built as a simple extension of the electroweak symmetry gauge group 
$SU_{L}(2)\otimes U_{Y}(1)$ of SM. It was first suggested in the work
of Voloshin~\cite{Vol}, who considered only the leptonic sector. Further developments of the model
including quarks and electroweak currents are presented in Ref.~\cite{341vicente}. The most attractive feature of this model
is the fact that the two chiralities of the lightest leptons are part of a fundamental representation of the gauge group $SU(4)_L$, unifying each lepton family in a single multiplet. Also, the explanation of the family replication comes from anomaly cancellation which, along with QCD asymptotic freedom, requires that  only three fermion families might be present in the spectrum, as in the $SU(3)_c\otimes SU(3)_L\otimes U(1)_N$ (331) models~\cite{331Pleitez}. Besides, new particles with distinct signature are predicted to appear close to the electroweak scale, like doubly charged vectors and scalars, with some of them carrying two units of lepton number (the so called bileptons), among other features also shared with the 331 models.

In the current
literature there are  341 models with or without 
exotic electric charges~\cite{Ponce,palcu}. Here we deal with the 341 model with exotic electric
charges, where we have reduced the scalar sector to only three scalar
quadruplets, which are enough to generate the exact number of Goldstone bosons needed to break the 341 symmetry to the $U(1)_{QED}$. In turn, we have disposed of one scalar quartet and the
scalar decuplet from the original version. These would be responsible for giving mass for the leptons and some of the quarks, but in our scheme they are not mandatory, since we are going to generate those masses by effective operators, as in Ref.~\cite{RM331}. Our approach poses a considerable reduction in the physical scalar spectrum, which is more suitable for
phenomenological studies. Moreover, we compute the running coupling of the abelian gauge group, $U(1)_X$, and show that there exists a Landau pole in this model, similar to the minimal 331 model~\cite{Dias,Dias2}. Nevertheless, in this work the 341 model breaks into a 331 model with right-handed ($331_{RHN}$)~\cite{341vicente}, or a 331 model with left-handed heavy neutral fermions ($331_{LHN}$)~\cite{331LHN}, and this means that such embedding of the lower symmetry into the 341 model implies a Landau pole for these 331 models too. The advantage of such embedding is that such models are known to possess cold dark matter candidates~\cite{331LHN,jcap331}, which are automatically incorporated into the body of this 341 model. Besides, any mechanism that employs non-renormalizable effective operators suppressed by a scale much higher than TeV is no longer an option in this scheme, which is the case of neutrino mass generation in some scenarios~\cite{331RHneutrinos}. Nonetheless,
we will show that all fermions, including neutrinos, gain mass from such operators even when the suppression scale is as low as the TeV scale~\footnote{Neutrino physics, including mass generation through effective operators, was investigated in the context of a different version of 341 model in Ref.~\cite{palcu}. There the author concludes that a cutoff scale around $10^{12}$~GeV was needed to implement his scheme, much higher than what we obtain for our model.}.

This work is organized as follows: section~\ref{sec1}, we introduce the 341
model with exotic electric charges. In section~\ref{sec2}, we discuss the perturbative limit of the model,
where we compute the value of its Landau pole, with the intent of
measuring the energy scale at which the model loses its perturbative
character. In section~\ref{sec3}, the fermion mass spectrum is obtained, including the neutrinos. Finally, in section~\ref{sec5}, we draw our conclusions.

\section{The 341 Model}
\label{sec1}
\indent The 341 model is a gauge extension of the electroweak gauge group of SM, that
possesses 12 gauge bosons besides the known 8 gluons of strong interactions and the electroweak, $W^\pm$, $Z^0$ and $\gamma$ of the SM. Since the strong gauge group $SU(3)_c$ is kept intact under the symmetry breakdown to $U(1)_{QED}$, we are going to simply omit it throughout this work.
In the following we present the field content of the 
341 model and write up its scalar potential and the Yukawa lagrangian, which are going to be relevant to our purposes in this work.

\subsection{Fermionic content}
In order to assign the fermionic content of the model, it is useful to recall that the electric charge operator is written as a linear combination of the diagonal generators of the electroweak gauge group,  $SU(4)_{L}\otimes U(1)_{X}$. This allows us to write the electric charge assignment for the fields in the fundamental representation of $SU(4)_{L}$ as,
{\small \begin{eqnarray}
\frac{Q}{e}&=& \frac{1}{2}\left( \lambda _{3}+\frac{1}{\sqrt{3}}b\lambda _{8}+
\frac{1}{\sqrt{6}}c\lambda _{15}\right) +X 
\nonumber \\
&=& \frac{1}{2}{\mbox diag} \left(1+\frac{b}{3}+\frac{c}{6}+2X,
-1+\frac{b}{3}+\frac{c}{6}+2X,
-\frac{2b}{3}+\frac{c}{6} +2X,
-\frac{c}{2}+2X
\right), \nonumber \\
\label{cargas1}
\end{eqnarray}}

\noindent and similarly for the anti-fundamental representation, exchanging the sign of the diagonal generators of $SU(4)$, $\lambda _{3}$, $\lambda _{8}$ and $\lambda _{15}$. Here, $b$ and $c$ are parameters to be fixed according to the field distribution in the quadruplets, and $X$ is the corresponding $U(1)_{X}$ charge.

Considering the SM content, which we should recover in the symmetry breakdown process,
the first two components of the left handed lepton quartet should recover the known doublet structure, meaning
that the first component is a neutrino and the second a negatively charged lepton. Moreover, as we are interested in accommodating the associated charge conjugate states for these leptons in the same multiplet,
recovering the $331_{RHN}$ model in the first symmetry breaking of 341, the third component of the quartet should be the charge conjugate of the neutrino and the fourth one should be the positive charged partner of the negative lepton, $\left( \nu
_{a}^{\prime },e_{a}^{\prime },\nu _{a}^{\prime c},e_{a}^{\prime c}\right)_L $. This association implies that $b=-1$, $c=-4$ and $X_{L_{aL}}=0$, which reduces the operator in Eq.~(\ref{cargas1}) to,
\begin{equation}
\frac{Q}{e}={\mbox diag}\left( 
X \,,\,\, 
X-1\,,\,\, 
X \,,\,\, 
X+1
\right)\,.
\end{equation}

Then, we can arrange all the leptons (including right-handed neutrinos) in quartets of $SU_{L}(4)$ symmetry as follows, 
\begin{equation}
L_{aL}=\left( 
\begin{array}{c}
\nu _{a} \\ 
e_{a} \\ 
\nu _{a}^{c} \\ 
e_{a}^{c}
\end{array}
\right) _{L}\sim \left( \mathbf{1},\mathbf{4},0\right) ,\text{ }
\end{equation}
where $a=$ $e,$ $\mu $, $\tau $ and the symbol $\sim $ indicates the respective representation
for the multiplet under the 341 symmetry.

For the first generation of quarks, we can assign the left-handed fields to the fundamental representation of the $SU_{L}(4)$ symmetry, while the right-handed ones transform trivially under this symmetry,
\begin{eqnarray}
Q_{1L} &=&\left( 
\begin{array}{c}
u_{1} \\ 
d_{1} \\ 
U_{1} \\ 
J_{1}
\end{array}
\right) _{L}\sim \left( \mathbf{3},\mathbf{4},\frac{2}{3}\right) ,\text{ } \\
u_{1R} &\sim &\left( \mathbf{3},\mathbf{1},\frac{2}{3}\right) ,\text{ }
d_{1R}\sim \left( \mathbf{3},\mathbf{1},-\frac{1}{3}\right) ,\text{ } \\
U_{1R} &\sim &\left( \mathbf{3},\mathbf{1},\frac{2}{3}\right) ,\text{ }
J_{1R}\sim \left( \mathbf{3},\mathbf{1},\frac{5}{3}\right) \,,
\end{eqnarray}
where $u_{1}$ and $d_{1}$ are the up and down quarks, respectively, $U$ and $J_{1}$ are
the new exotic quarks with electric charges, $2/3$ and $5/3$, respectively.

The second and third families of left-handed quarks are arranged in the anti-quartet representation of the $SU_{L}(4)$ symmetry, and right-handed quarks in the singlet representation,
\begin{eqnarray}
Q_{iL} &=&\left( 
\begin{array}{c}
d_{i} \\ 
u_{i} \\ 
D_{i} \\ 
J_{i}
\end{array}
\right) _{L}\sim \left( \mathbf{3},\mathbf{4}^{\ast },-\frac{1}{3}\right) ,
\text{ } \\
u_{iR} &\sim &\left( \mathbf{3},\mathbf{1},\frac{2}{3}\right) ,\text{ }
d_{iR}\sim \left( \mathbf{3},\mathbf{1},-\frac{1}{3}\right) ,\text{ } \\
D_{iR}\text{ } &\sim &\left( \mathbf{3},\mathbf{1},-\frac{1}{3}\right) ,
\text{ }J_{iR}\sim \left( \mathbf{3},\mathbf{1},-\frac{4}{3}
\right) ,
\end{eqnarray}
where $i=2,3.$ $D_{i}$ and $J_{i\text{ }}$ are exotic quarks with
electric charges $-\frac{1}{3}$ and $-\frac{4}{3}$, respectively. This arrangement 
in different representations for the quark families is a
consistency requirement such that the model be free of anomalies. Next we explore the scalar content of the model.

\subsection{Scalar content}
\indent
Originally~\cite{341vicente}, the 341 model was built with four scalar quartets in order to engender the symmetry breakdown to $U(1)_{QED}$, give mass to all quarks and avoid mixing among ordinary and exotic quarks. Also, one scalar decuplet was introduced so as to generate charged lepton masses. One of the purposes of this work is to show that we can diminish the scalar content and still have all desirable properties concerning the symmetry breakdown and the fermion masses (including neutrinos). Here we present the minimal scalar content which will allow us to accomplish this goal,
\begin{eqnarray}
\eta &=&\left( 
\begin{array}{c}
\eta _{1}^{0} \\ 
\eta _{1}^{-} \\ 
\eta _{2}^{0} \\ 
\eta _{2}^{+}
\end{array}
\right)\,,
\rho =\left( 
\begin{array}{c}
\rho _{1}^{+} \\ 
\rho ^{0} \\ 
\rho _{2}^{+} \\ 
\rho ^{++}
\end{array}
\right) \,,
\chi =\left( 
\begin{array}{c}
\chi _{1}^{-} \\ 
\chi ^{--} \\ 
\chi _{2}^{-} \\ 
\chi ^{0}
\end{array}
\right) \,,
\end{eqnarray}
which transform as, $\eta  \sim \left( \mathbf{1},\mathbf{4},0\right)$, $\rho \sim \left( \mathbf{1},\mathbf{4},1\right)$ and $\chi \sim \left( \mathbf{1},\mathbf{4},-1\right)$.
The desired pattern of symmetry breakdown can be obtained if we assume that the following neutral components develop nontrivial vacuum expectation value (VEV), $\left\langle \chi ^{0}\right\rangle =\frac{1}{\sqrt{2}}v_{\chi
}$, $\left\langle \eta _{2}^{0}\right\rangle =\frac{1}{\sqrt{2}}v_{\eta }$ and $\left\langle \rho ^{0}\right\rangle =\frac{1}{\sqrt{2}}v_{\rho }$. The VEV $v_{\chi}$ is responsible for the first step in breaking the 341 symmetry to 331, while $v_\eta$ breaks the 331 symmetry to 321, the SM, and the final breaking to $U(1)_{QED}$ is provided by $v_\rho$. In this way we can impose, $v_\chi > v_\eta > v_\rho = 246$~GeV. Indeed, as we are going to show later, $v_\chi$ and $v_\eta$ should be no more than few TeV or we lose perturbativity, which makes this model interesting in the sense it can be promptly tested at LHC or discarded soon.

The reason we choose the $\eta $ quadruplet to develop VEV only in the third component is related to the fact that we do not want mixing among ordinary and exotic quarks on the Yukawa lagrangian, which guarantees the usual CKM mixing in the quark sector. Of course this is only possible if we also add a new symmetry to the model, in our case this is going to be a $Z_3$ discrete symmetry that we implement later on. 

With the scalar and fermionic content defined, we can write down the lagrangian of the model, invariant under the gauge symmetry and the additional discrete symmetry mentioned above. Next, we add this symmetry and present the Yukawa and scalar Lagrangian of the model, also defining our gauge boson spectrum.

\subsection{The $Z_3$ symmetry and 341 lagrangian}
\indent Let us impose a discrete $Z_3$ symmetry that will be suitable to avoid the mixing among ordinary and exotic quarks but also allow for an appropriate scenario for generating fermion masses through effective operators, as we will see later. The  $Z_{3}$ charges are $(e\,,$ $\omega\,,$ $\omega ^{2})$,
with $\omega ^{3}\equiv e$, where $e$ is the identity element of the group. We then assign $Z_3$ charges to the 341 model fields given in table~\ref{table1}.
\begin{table}[htb]
\centering
\begin{tabular}{||l|l|l|l|l|l|l||}
\hline
$L_L$ & $u_{a_R}$ & $d_{a_R}$  & $J_{1_R}$ & $J_{i_R}$ & $\rho$ & $\chi$ \\ \hline
$\,\,\omega$ & $\,\,\omega$ & $\,\,\omega^2$  & $\,\,\omega^2$ & $\,\,\omega$ & $\,\,\omega$ & $\,\,\omega$
\\ \hline
\end{tabular}%
\caption{$Z_3$ symmetry transformation properties for the 341 fields. Those fields not present in this table transform trivially under this discrete group.}
\label{table1}
\end{table}

The Yukawa lagrangian, ${\cal L}_{Y}$, invariant under this $Z_{3}$ and gauge 
symmetry, is given by,
\begin{eqnarray}
{\cal L}_Y &=&\lambda _{11}^{J}\overline{Q_{1_{L}}}\chi
J_{1_{R}}+\lambda _{ij}^{J^{ }}\overline{Q_{i_{L}}}\chi ^{\ast
}J_{j_{R}}^{ }+\lambda _{1a}^{d}\overline{Q_{1_{L}}}\rho d_{a_{R}} 
\nonumber \\
&&+\lambda _{ia}^{u}\overline{Q_{i_{L}}}\rho ^{\ast }u_{a_{R}}+\lambda
_{11}^{U}\overline{Q_{1_{L}}}\eta U_{1_{R}}+\lambda _{ij}^{D}\overline{
Q_{i_{L}}}\eta ^{\ast }D_{j_{R}}+h.c.,
\label{yukawa}
\end{eqnarray}
where $\lambda _{11}^{J},$ $\lambda _{ij}^{J},$ $\lambda _{1a}^{d},$ $
\lambda _{ia}^{u},$ $\lambda _{11}^{U},$ $\lambda _{ij}^{D}$ are the Yukawa
coupling constants ($a=1,2,3$ and $i,j=2,3)$.

${\cal L}_Y$ only generates some of the quark masses, leaving the leptons massless. Namely, the exotic quarks, $J_{a}$ ($a=1,2,3$), through their coupling to $\chi$ gain mass proportional to $v_\chi$, while the exotic,  $U_{1}$, $D_2$ and $D_3$, quarks receive mass from $v_\eta$. The second and third families of ordinary up-type quarks and first family of down-type quarks gain mass proportional to $v_\rho$. Besides this incomplete fermion mass spectrum, the Yukawa lagrangian above is unable to explain the observed mixing in the quark sector, parametrized by the CKM matrix. Curiously, as we will see later, the model has a peculiar characteristic that allows it to provide the remaining masses and mixings through effective operators, so that the issues pointed here are not going to jeopardize our scheme of reducing the scalar spectrum.

The most general scalar potential, respecting the imposed symmetries for the 341 model here studied is~\footnote{If it were not for the $Z_3$ symmetry, an additional term would remain in the scalar potential, namely, $\left[ \lambda _{10}\left( \eta ^{\dag }\chi \right) \left(
\eta ^{\dag }\rho \right) +h.c.\right]$.},
\begin{eqnarray}
V\left( \eta ,\rho ,\chi \right) &=&\mu _\eta^{2}\eta ^{\dag }\eta +\mu
_\rho^{2}\rho ^{\dag }\rho +\mu _\chi^{2}\chi ^{\dag }\chi +\lambda _{1}\left(
\eta ^{\dag }\eta \right) ^{2}+\lambda _{2}\left( \rho ^{\dag }\rho \right)
^{2}+\lambda _{3}\left( \chi ^{\dag }\chi \right) ^{2}  \nonumber \\
&&+ \lambda _{4}\left( \eta ^{\dag }\eta \right)\left( \rho ^{\dag
}\rho \right) +\lambda _{5}\left( \eta ^{\dag }\eta \right)\left( \chi ^{\dag }\chi \right) +\lambda
_{6}\left( \rho ^{\dag }\rho \right) \left( \chi ^{\dag }\chi \right)
\nonumber \\
&&+\lambda _{7}\left( \rho ^{\dag }\eta \right) \left( \eta ^{\dag }\rho
\right)  +\lambda _{8}\left( \chi ^{\dag }\eta \right) \left( \eta ^{\dag }\chi
\right) +\lambda _{9}\left( \rho ^{\dag }\chi \right) \left( \chi ^{\dag
}\rho \right)\,,  
\label{pot}
\end{eqnarray}
where $\mu_{\eta,\rho,\chi}^{2}$ are the mass dimension parameters and the $\lambda$s are dimensionless coupling constants. Observe that, due to the absence of a fourth scalar quartet, there is no anti-symmetric quartic coupling as in the original version of the model~\cite{341vicente}. This scalar potential, after spontaneous symmetry breaking is driven by the assumed VEVs of neutral scalars, produces the right amount of Goldstone bosons which give  masses to all gauge bosons, but the photon. Since the scalar and vector bosons spectrum is not going to be explored in this work, except for the fact that we are going to count their degrees of freedom contributing to the running couplings in the next section, we expose their mass eigenvalues and eigenstates in the appendix. Also, since we have to correctly count the fields degrees of freedom that are not decoupled at each stage in the running, it is good to have it clear which are the effective degrees of freedom by identifying their transformation properties 
after each step in the symmetry breakdown. We present that in the appendix too. Next we address the question of the existence of a Landau pole for this model.

\section{341 model Landau pole}
\label{sec2}
\indent In this section we obtain the first important result of this work. It concerns the existence of a Landau pole for the 341 model we are dealing with. The existence of such a pole would imply a limit scale, a natural cutoff for the model, where it loses its perturbative character. We are motivated by the results obtained in Ref.~\cite{Dias}, where the minimal 331 model is shown to possess a Landau pole at few TeV scale, firstly guessed because there is a relation between the gauge couplings where a pole appears as the electroweak mixing angle, $\sin^2\theta_W\equiv s^2_W$, approaches 1/4, and confirmed through the renormalization group equations (RGE) for the $U(1)_N$ gauge coupling (the abelian group in the minimal 331 model). Curiously, this does not happen to another version of the 331 model, namely, the $331_{RHN}$ model. However, if there is a Landau pole for our 341 model, we are automatically showing that the $331_{RHN}$ model embedded in it is also limited by this pole, meaning that some care 
has to be taken when dealing with effective operators in this model. Also, this result constrains the maximal mass scale available for its particle spectrum which, from the phenomenological point of view, are bound to be found or discarded in the LHC and/or next generation of colliders.

The presence of a Landau pole in this model can be asserted from observing that the relation between the gauge coupling constants is,
\begin{equation}
\frac{g_{X}^{2}}{g_{L}^{2}}=\frac{s^2_W}{1-4s^2_W}\,.
\label{LPsw}
\end{equation}
Such a relation is also present in the minimal 331 model~\cite{331Pleitez} and indicates that the $s^2_W$ is limited to  $s^2_W= 1\slash 4$. Of course, when $s^2_W$ approaches this value, the abelian gauge coupling goes to infinity, $g_X\rightarrow \infty$ (as far as $g_L$ remains finite), signaling the presence of a Landau pole and showing that the theory has lost its perturbative limit, much before that~\cite{Dias}. 

In order to derive the Landau pole for the 341 model, we first recall that the evolution of the couplings according to the RGE is,
\begin{equation}
\frac{1}{\alpha_i(\mu)}=\frac{1}{\alpha_i(\mu_0)}+\frac{1}{2\pi}b_i\ln(\frac{\mu_0}{\mu})\,,
\label{RGEs}
\end{equation}
where the index $i$ labels the associated gauge group coupling and, in a
generic unitary gauge group, the coefficients $b_{i}$ at one-loop are given by,
\begin{equation}
b_{i}=\frac{2}{3}{\sum_{Fermions}}T_{R}\left( F\right) _{i}+\frac{1
}{3}{\sum_{Scalars}}T_{R}\left( S\right) _{i}-\frac{11}{3}
C_{2}\left( G\right) _{i}\,  
\label{b}
\end{equation}
In the above expression, the Dynkin index, $T_{R}\left( I\right)$, is defined through the trace $Tr\left[ T^{i}\left(
I\right) T^{j}\left( I\right) \right] = \-T_{R}\left( I\right) \delta ^{ij}$, with $T^{i}$ being the generators in some representation $R$ of the gauge group in question, and $I=S\,,F$ represent complex scalars and Weyl fermions, respectively. $C_{2}\left( G\right) $ is the quadratic Casimir operator for the adjoint representation of the gauge group $G$. For the fundamental representation, $T_{R}\left( I\right) =1/2$ and $C_{2}\left( G\right) =N$ for $SU(N)$, while for the abelian gauge group,  $U\left( 1\right) _{y}$, we have $C_{2}\left(
G\right) =0$ and use 
$\sum T_{R}\left( F,S\right) =\sum y^{2}$ where $y=Y/2$ for the SM and $y=X$
for the 341 model~\cite{Jones}.

From now on, the coupling constants for the $341$ model are denoted by $
\alpha _{3}$, $\alpha _{4L}$ and$\ \alpha _{X}$ relative to the groups $
SU(3)_{c}$, $SU(4)_{L}$ and $U(1)_{X}$, respectively, where we have written $\alpha
_{i}=g_{i}^2/4\pi $. Analogously, the coupling constants for the $331$ model are denoted by $\alpha _{3}$, $\alpha _{3L}$,  $\alpha _{N}$ and, for the $321$ SM,
they are $\alpha _{3}$, $\alpha _{2}$ and$\ \alpha _{1}$, respectively. Observe that, since we are assuming the embedding $SU(4)_L\supset SU(3)_L\supset SU(2)_L$, their respective couplings are the same, implying $\alpha _{4L}(\mu_{341})=\alpha _{3L}(\mu_{341})$ and $\alpha_{3L}(\mu_{331})=\alpha _{2}(\mu_{331})$.

We should proceed step by step in order to obtain the characteristic scale for which the 341 model loses its perturbative behavior and also to determine the Landau pole. Starting with the running of the SM gauge couplings below the 331 model breaking scale $\mu_{331}\equiv v_\eta/\sqrt{2}$, where the contributing degrees of freedom are those related to the fields which acquire mass below $\mu_{331}$. In this reduced model, there is no new contribution to the SM running couplings besides those belonging to the SM alone (see the fields decomposition in the appendix). We then have,
\begin{equation}
\frac{1}{\alpha _{i}\left( \mu \right) }=\frac{1}{\alpha _{i}\left(
M_{Z}\right) }+\frac{1}{2\pi }b_{i}\ln \left( \frac{M_{Z}}{\mu }\right)\,,
\text{ for }\mu \leq \mu _{331}.  
\label{rge1}
\end{equation}
From this equation we can easily compute $\alpha_1$ and $\alpha_2$ at any scale $\mu$ since we can relate these couplings to the electromagnetic coupling $\alpha$ and $s^2_W$ at the $M_Z$ scale, $\alpha _{1}\left( M_{Z}\right) =\alpha \left( M_{Z}\right)
/c_{W}^{2}\left( M_{Z}\right) $ and $\alpha _{2}\left( M_{Z}\right) =\alpha
\left( M_{Z}\right) /s^2_W\left( M_{Z}\right)$. However, we are interested in obtaining a relation among the SM and the 331 model couplings, in order to finally get to the 341 model couplings. In particular, we wish to know the running of $\alpha_X$, which is the coupling that fastly increases as we approach $s_W^2=1/4$. To do so, we first write $\alpha_N(\mu)$,
\begin{equation}
\frac{1}{\alpha _{N}\left( \mu \right) }=\frac{1}{\alpha _{N}\left( \mu
_{331}\right) }+\frac{1}{2\pi }b_{N}\ln \left( \frac{\mu _{331}}{\mu }%
\right) ,\text{ for }\mu \geq \mu _{331}\,. 
\label{rge0}
\end{equation}
This can be related to the SM couplings in Eq.~(\ref{rge1}) by noticing the following expressions for the running of $s^2_W$,
\begin{equation}
s^2_W\left( \mu \right) =\frac{1}{1+\alpha _{2}\left( \mu \right)
/\alpha _{1}\left( \mu \right) } \text{ for }\mu \leq \mu _{331}\,,
\end{equation}
and 
\begin{equation}
s^2_W\left( \mu \right) =\frac{3\alpha _{N}\left( \mu \right) }{3\alpha
_{3L}\left( \mu \right) +4\alpha _{N}\left( \mu \right) } \text{ for }
\mu _{331} \leq \mu \leq \mu _{341}\,,
\label{sw331}
\end{equation}
that should match at the scale $\mu_{331}$, yielding,
\begin{equation}
\frac{1}{\alpha _{N}\left( \mu _{331}\right) }=\frac{1}{\alpha _{1}\left(
\mu _{331}\right) }-\frac{1}{3}\frac{1}{\alpha _{2}\left( \mu _{331}\right) }\,,
\label{rge1.0}
\end{equation}
remembering that $\alpha _{2}\left( \mu _{331}\right) =\alpha _{3L}\left(
\mu _{331}\right)$. 

For the sake of completeness, when we reach the $\mu_{341}$ scale, the content of 341 model becomes relevant (at least the light modes) and the mixing angle evolves as dictated by Eq.~(\ref{LPsw}), which can be inverted to give,
\begin{equation}
s^2_W\left( \mu \right) =\frac{1}{4}\frac{1}{(1 +\alpha _{4L}\left( \mu \right)/4\alpha _{X}\left( \mu \right)  )} \text{ for }
 \mu _{341} \geq \mu \,.
\label{sw341}
\end{equation}
Notice that this expression is exactly the same as Eq.~(12) in Ref.~\cite{Dias}, obtained in the context of the minimal 331 model. 

Then, we substitute 
Eq.~(\ref{rge1}) for $\alpha_1(\mu_{331})$ and $\alpha_2(\mu_{331})$ into Eq.~(\ref{rge1.0}), and plug the result in Eq.~(\ref{rge0}), to obtain,
\begin{equation}
\frac{1}{\alpha _{N}\left( \mu \right) }=\frac{3-4s^2_W\left(
M_{Z}\right) }{3\alpha \left( M_{Z}\right) }+\frac{1}{2\pi }\left( b_{1}-
\frac{1}{3}b_{2}\right) \ln \left( \frac{M_{Z}}{\mu _{331}}\right) +\frac{1}{
2\pi }b_{N}\ln \left( \frac{\mu _{331}}{\mu }\right)\,,
\label{RGEaN}
\end{equation}
valid for $\mu \geq \mu _{331}$. Here we are going to use $s^2_W\left( M_{Z}\right) =0.2311$, $\alpha \left( M_{Z}\right) =1/128$ and $M_{Z}=91.188$~GeV~\cite{Naka}. The RGE coefficients $b_1$ and $b_2$ are computed for the SM content only, while $b_N$ receives contribution from all SM fields plus those fields whose masses are below the $\mu_{341}$ scale, which means that we include all the 331 fields, leading to,
\begin{equation}
\left(b_1,\,b_2,\,b_3\right) = \left(\frac{41}{6},\,-\frac{19}{6},\,-7 \right) \text{   for the SM }\,,
\label{biSM}
\end{equation}
\begin{equation}
\left(b_N,\,b_{3L},\,b_3\right) = \left(\frac{77}{9},\,-\frac{20}{3},\,-5 \right) \text{   for the 331 }\,.
\label{bi331}
\end{equation}
where we have included the $b_3$ coefficient for the color gauge group just for completeness.

Now, we can proceed to compute the perturbative limit  imposed by the presence of a Landau pole to the $U(1)_{X}$ gauge coupling, $\alpha _{X}$. The associated evolution equation is,
\begin{equation}
\frac{1}{\alpha _{X}\left( \mu \right) }=\frac{1}{\alpha _{X}\left( \mu
_{341}\right) }+\frac{1}{2\pi }b_{X}\ln \left( \frac{\mu _{341}}{\mu }
\right) ,\text{ for }\mu \geq \mu _{341}\,,  
\label{rge2}
\end{equation}
%
%
%
%
%
%
where,
\begin{equation}
\frac{1}{\alpha _{X}\left( \mu _{341}\right) }=\frac{1}{\alpha _{N}\left(
\mu _{341}\right) }-\frac{8}{3}\frac{1}{\alpha _{4L}\left( \mu _{341}\right) 
}\,.
\label{rge3}
\end{equation}
The values of the couplings on the r.h.s. of Eq.~(\ref{rge3}) can be easily computed from Eqs.~(\ref{RGEs}) and (\ref{RGEaN}) and considering the fact that  $\alpha _{3L}\left( \mu _{341}\right) =\alpha _{4L}\left(
\mu _{341}\right) $.
%
%
%
%
%
%
The RGE coefficients used in this equation were computed by considering all the 341 model fields,
and are given by,
\begin{equation}
\left(b_X,\,b_{4L},\,b_3\right) = \left(\frac{76}{3},\,-\frac{61}{6},\,-3 \right)\,.
\label{bi341}
\end{equation}

We are now able to find the value for the Landau pole by using Eq.~(\ref{rge2}). We impose the $\lim_{\mu\to\Lambda}\alpha_X(\mu)=\infty$ in that equation and then, inverting for the scale $\Lambda$, one obtains
\begin{equation}
\Lambda =\mu _{341}\exp \left( \frac{2\pi }{b_{X}\text{ }\alpha _{X}\left(
\mu _{341}\right) }\right)\,.
\label{lambda}
\end{equation}

Another important scale that we have to be concerned with is related to the point where the model loses its perturbative character. We define this scale as the point where $\alpha_X(\Lambda^\prime)=1$, and from Eq.~(\ref{rge2}) it can be written as,
\begin{equation}
\Lambda^{\prime }=\mu _{341}\exp \left\{ \left( \frac{2\pi }{b_{X}}\right) \left[
\frac{1}{\alpha _{X}\left( \mu _{341}\right) }-1\right] \right\}\,,
\label{mlinha}
\end{equation}
observing that self consistency requires that $\Lambda^\prime < \Lambda$.

\begin{figure}
\centering
\includegraphics[scale=0.8]{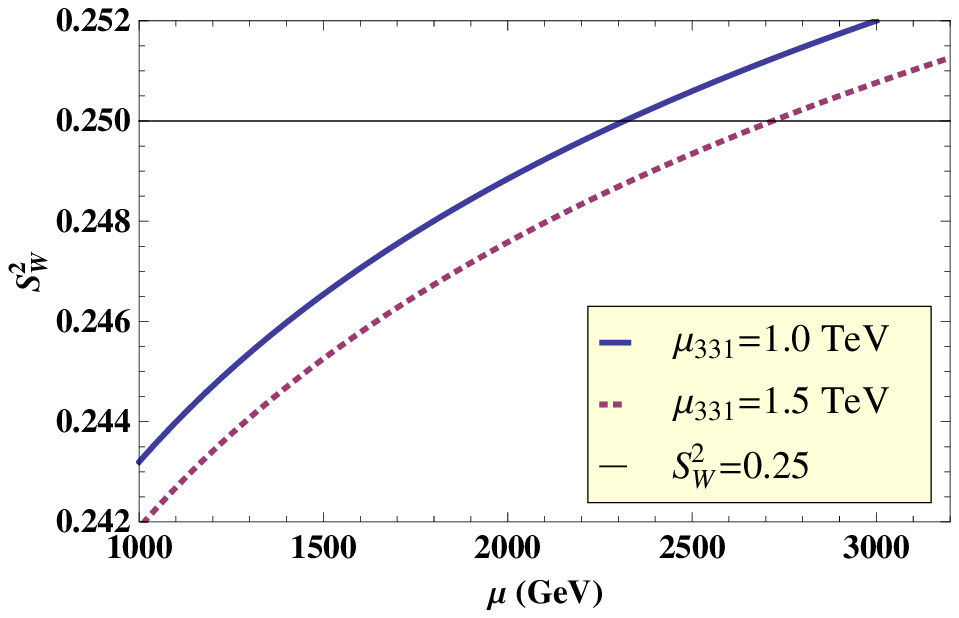}
\includegraphics[scale=0.8]{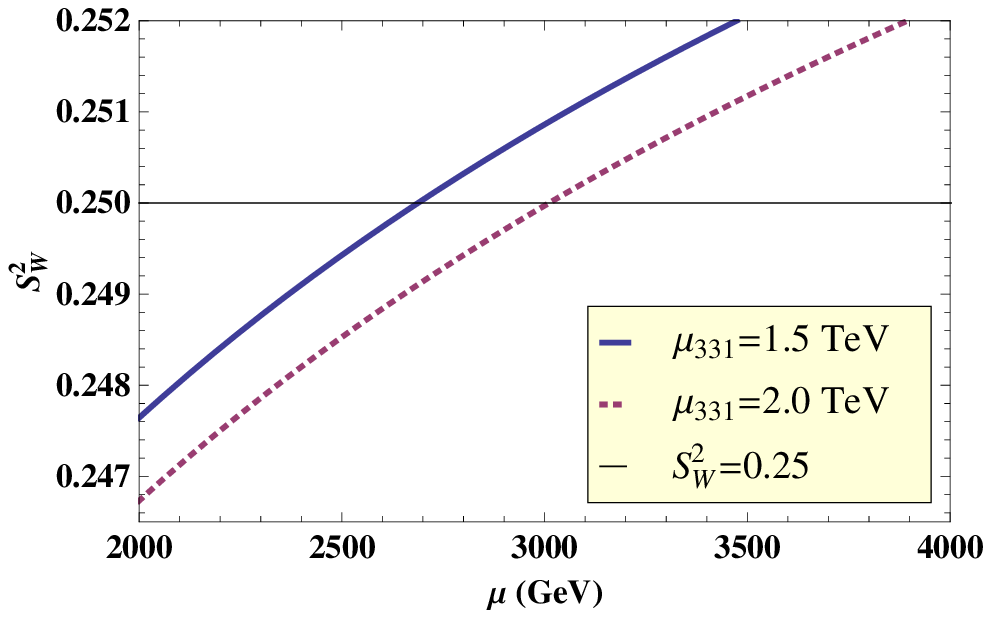}
\caption{Running of the electroweak mixing angle, $s_W^2(\mu)$ as a function of energy scale $\mu$ for $\mu_{341} = 1.5$~TeV (top panel) and $\mu_{341} = 2$~TeV (bottom panel). Here all the content of the 341 model is considered.}
\label{fig1}
\end{figure}

\begin{figure}
\centering
\includegraphics[scale=0.8]{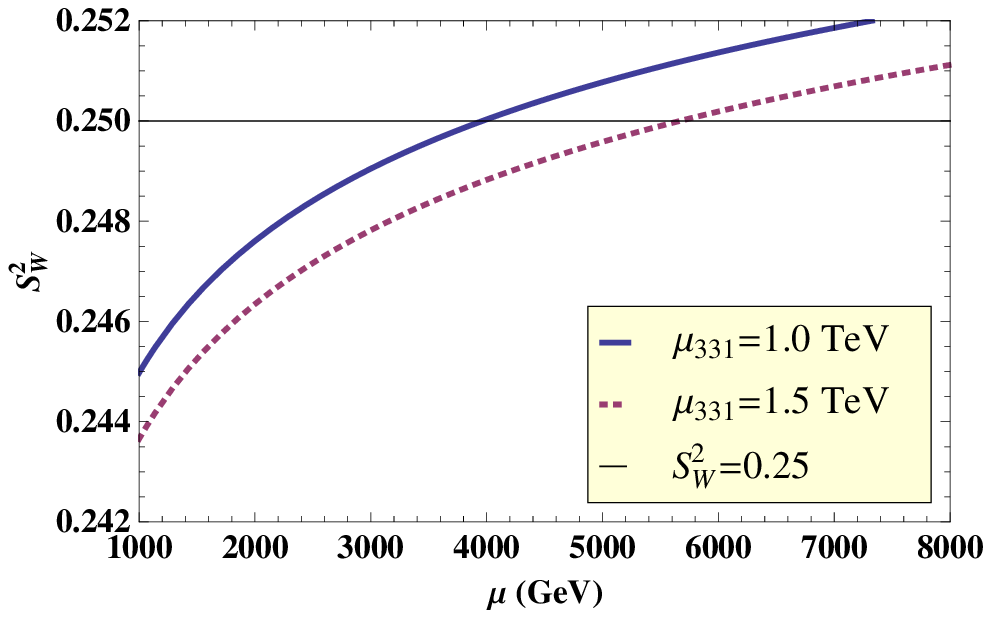}
\includegraphics[scale=0.8]{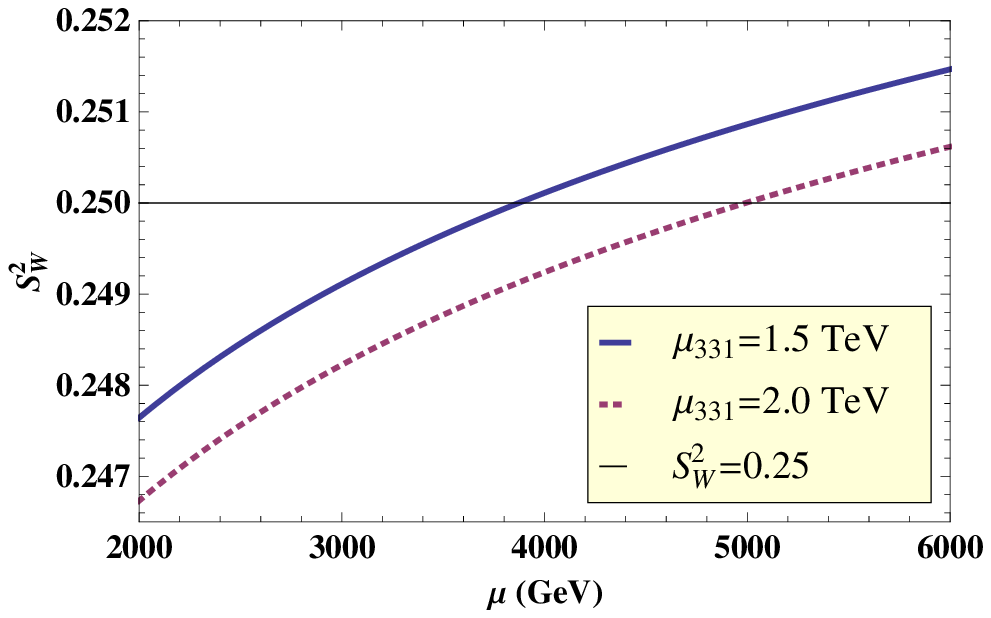}
\caption{The same as Fig.~\ref{fig1}, but here the exotic quarks typical of 341 model and which are singlets under 331 symmetry are decoupled. }
\label{fig2}
\end{figure}

We show in Fig.~\ref{fig1} the running of the $s^2_W(\mu)$ for two values of the 341 scale, $\mu_{341} = 1.5$~TeV (left panel) and $\mu_{341} = 2$~TeV (right panel), including all 341 content, while Fig.~\ref{fig2} the same situations are depicted but the typical 341 exotic quarks (which are singlets under 331 symmetry) are decoupled.
We see from Fig.~\ref{fig1}, where all 341 particles are taken into account, that the Landau pole scale $\Lambda$, obtained when $s_W^2\rightarrow 1\slash 4$, is larger for the case where the 341 and 331 scales approach each other. As a matter of fact, we could admit that 341 model breaks directly to the SM gauge group, which would be equivalent to make the reasonable assumption that $\mu_{341} = \mu_{331}$ (or $v_\chi = v_\eta$). Also, the larger the $\mu_{341}$ scale is, the larger the Landau pole scale for the case where all the 341 model particle content is taken into account.  Nevertheless, although the same qualitative profile is obtained when the 341 exotic quarks are decoupled, the larger Landau pole is for a lower $\mu_{341}$ scale, namely $\mu_{341} = 1.5$~TeV. 

\begin{figure}
\centering
\includegraphics[scale=0.8]{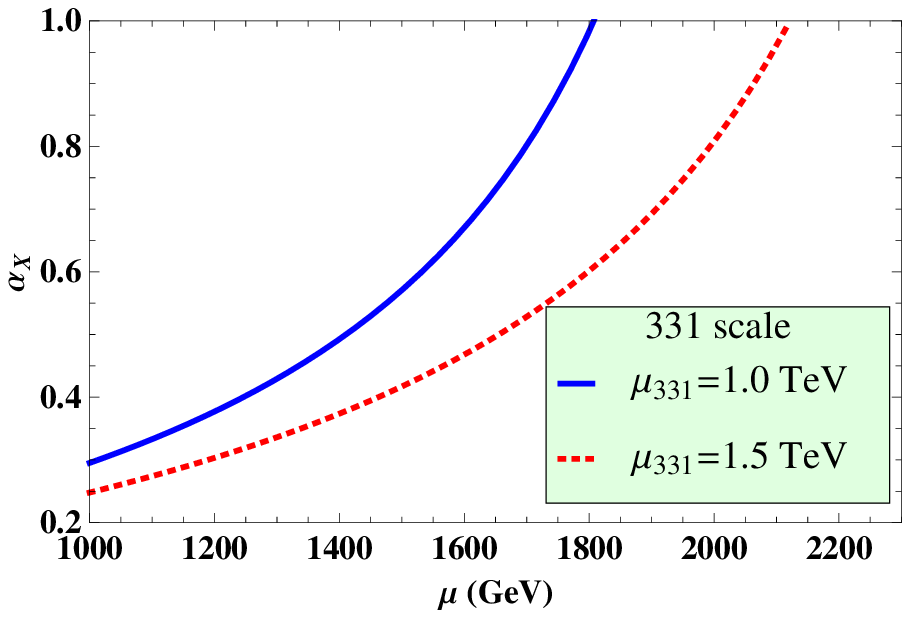}
\includegraphics[scale=0.8]{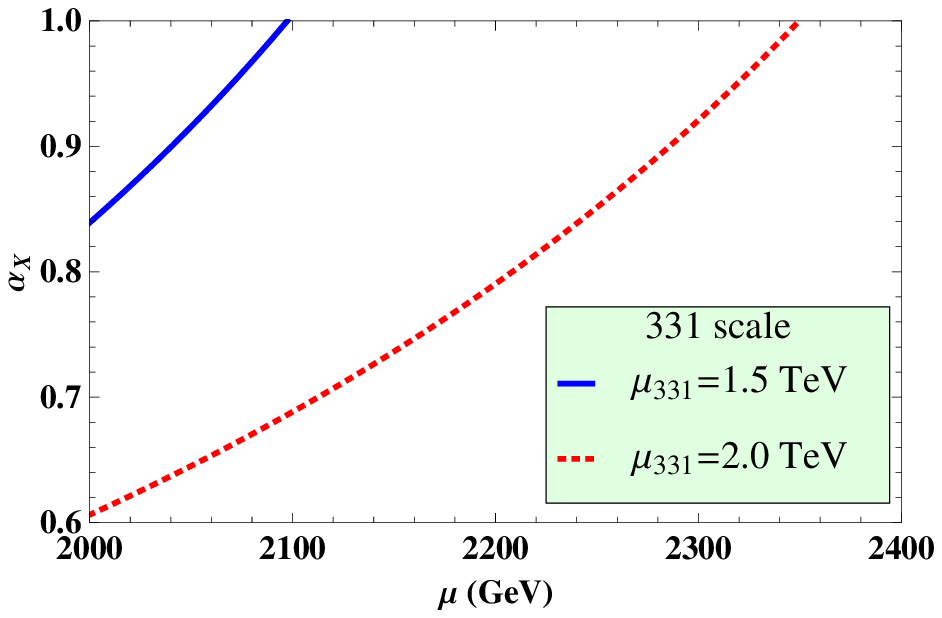}
\caption{$U(1)_X$ running coupling, $\alpha_X(\mu)$ as a function of energy scale $\mu$ for $\mu_{341} = 1.5$~TeV (top panel) and $\mu_{341} = 2$~TeV (bottom panel). Here all the content of the 341 model is considered.}
\label{fig3}
\end{figure}

\begin{figure}
\centering
\includegraphics[scale=0.8]{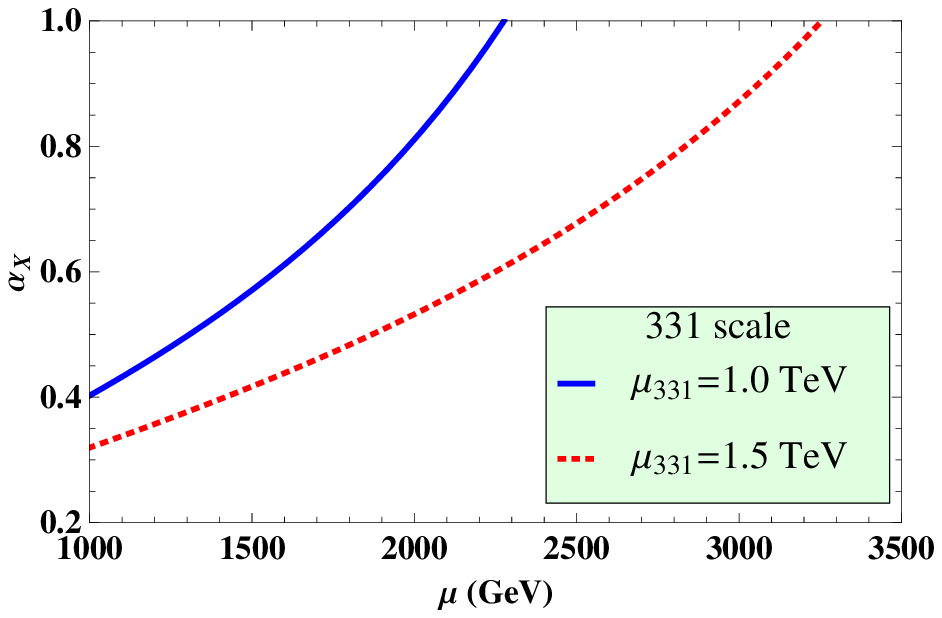}
\includegraphics[scale=0.8]{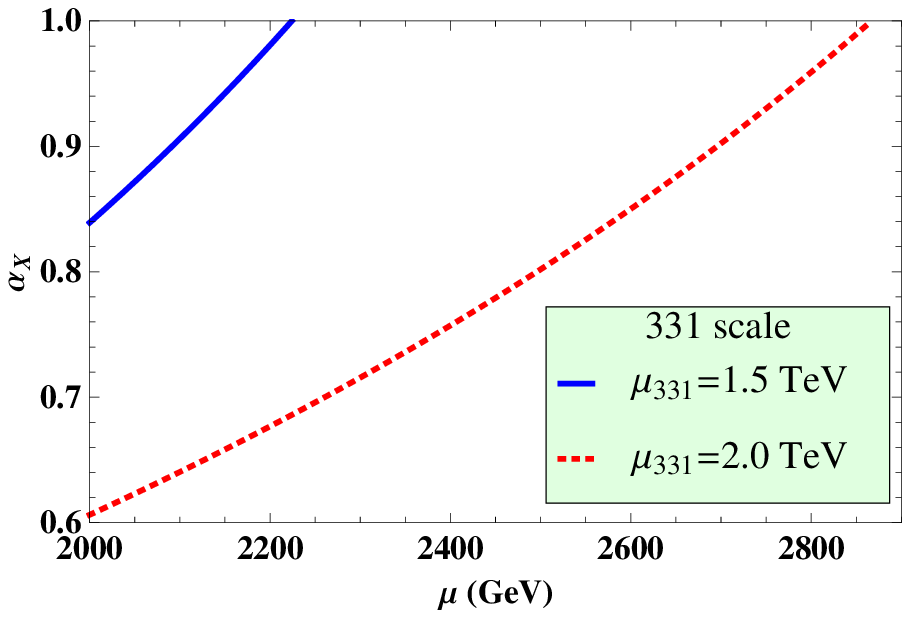}
\caption{The same as Fig.~\ref{fig3}, but here the exotic quarks typical of 341 model and which are singlets under 331 symmetry are decoupled. }
\label{fig4}
\end{figure}

In Fig.~\ref{fig3} we show the plot for the $U(1)_X$ running coupling for two values of the 341 scale, $\mu_{341} = 1.5$~TeV (left panel) and $\mu_{341} = 2$~TeV (right panel), including all 341 content. In Fig.~\ref{fig4} we plot the same situations but with  the typical 341 exotic quarks decoupled.
From these figures we can check that the perturbative scale $\Lambda^\prime$ is larger when the 331 and 341 scales approach each other. But they are larger for a larger $\mu_{341}$ scale in the case where all 341 particles are considered, but in the case where the exotic 341 quarks are decoupled, the larger $\Lambda^\prime$ is obtained when $\mu_{341}$ is smaller. The results presented in these plots show that there is no significant difference in the running of  $s_W^2(\mu)$  and  $\alpha_X(\mu)$, when $\mu_{331}$ scale approaches the $\mu_{341}$ scale, and this legitimates the simplifying assumption that $v_\eta\approx v_\chi$, that we use to obtain the whole particles' mass spectrum.

Below we show the numerical values for $\Lambda$ and $\Lambda^\prime$ considering
$\mu _{341} = 2.0$~TeV and varying the $\mu _{331}$ scale (all scales in units of TeV):
\begin{table}[htb]
\centering
\begin{tabular}{||l||l||l||}
\hline
$\mu _{331}$ & $\Lambda^{\prime }$ & $\Lambda$ \\ \hline
$1.5$ & $2.09$ $\left( 2.22\right) $ & $2.68$ $\left( 3.87\right) $
\\ \hline
$1.6$ & $2.15$ $\left(2.35\right) $ & $2.76$ $\left( 4.09\right) $
\\ \hline
$1.7$ & $2.20$ $\left( 2.48\right) $ & $2.82$ $\left( 4.32\right) $
\\ \hline
$1.8$ & $2.25$ $\left( 2.61\right) $ & $2.88$ $\left( 4.54\right) $
\\ \hline
$1.9$ & $2.30$ $\left( 2.74\right) $ & $2.95$ $\left( 4.77\right) $
\\ \hline
$2.0$ & $2.35$ $\left( 2.86\right) $ & $3.01$ $\left( 4.99\right) $
\\ \hline
\end{tabular}%
\caption{The scales $\Lambda^\prime$ (indicating loss of perturbativity) and $\Lambda$ (the Landau pole) for $\mu _{341}=2.0$~TeV and varying $\mu _{331}$. The number within paratheses refer to the case where the exotic quarks from the 341 model are decoupled. All results in TeV units.}
\label{tab:tab1}
\end{table}

In table~\ref{tab:tab1} we have also shown the values of $\Lambda$ and $\Lambda^\prime$ when the exotic quarks, $J_a$ (with $a=1,\,2,\,3$), are decoupled, in which case we used the new value for the RGE abelian coefficient in Eq.~(\ref{bi341}), $b_X = 34/3$. We observe that the highest values for the Landau pole scale are reached as the 331 symmetry breaking scale approaches that of 341 model, and can be as high as $\Lambda\simeq 5$~TeV when the exotic quarks are decoupled, although perturbativity is lost about half that scale. We do not go beyond $\mu_{341}=2$~TeV because our results start becoming inconsistent in the sense that $\Lambda^\prime$ gets smaller than $\mu_{341}$, in other words, we lose perturbativity at lower energies than 341 symmetry breaking scale. This has a deep implication for such class of models, imposing that they must manifest somewhere close to the electroweak scale and may be promptly tested at LHC and/or ILC.

Finally, the economical 331 model~\cite{ecolong}, along with an extra scalar triplet and three singlet scalars, can be embedded in the model here studied. The same is true for the 331 model when the right-handed neutrino is exchanged by a new neutral fermion, $N_L$~\cite{331LHN}.
This may have interesting implications for these models as well as for the 341 model developed here. For the 331 models, it is automatic that they are not valid up to some arbitrary high energy scale, being limited to the Landau pole of our 341 model, $\Lambda\approx 2-5$~TeV. Such a low cutoff scale is an obstacle for some neutrino mass generation mechanisms in these models~\cite{331RHneutrinos}, a problem we address here with an appropriate discrete symmetry and non renormalizable effective operators. In what concerns the advantages for the 341 model, we can mention the fact that some of these models possess natural dark matter candidates~\cite{331LHN,jcap331}, which can be explored in the context of our 341 model. Besides, since the scalar resonance discovered at LHC~\cite{LHC} can be perfectly adjusted into the 331 models~\cite{331higgs}, its phenomenology can be recovered by our model as well, with some room for extra new gauge bosons and scalars to be tested at LHC. Besides, it is possible that the 
new neutral singlet fields (under the 331 symmetry), may increase the number of dark matter candidates, a question to be investigated elsewhere.

Next we show that the fermion mass spectrum can be reproduced by means of non renormalizable effective operators, once the $Z_3$ symmetry introduced in the last section can be implemented in the model.

\section{Fermion Masses}
\label{sec3}
\indent 
Now that we have established the highest energy scale that works as a cutoff for this 341 model, we can make use of the $Z_3$ symmetry imposed through the assignments in Table~\ref{table1} in order to generate the observed fermion mass spectrum. In the original version of 341 model, this spectrum was obtained by means of four scalar quartets plus a decuplet. The presence of a natural cutoff scale, $\Lambda\simeq 5$~TeV, due to the existence of a Landau pole at that scale, turns this scenario suitable for introducing non renormalizable effective operators to get realistic fermion masses and mixing without plaguing the model with too many scalar multiplets.

Some of the SM quarks get contributions to their masses from the following dimension-6 effective operators: 
\begin{eqnarray}
&&\frac{\lambda
_{1a}^{u}\text{ }}{\Lambda ^{2}}\varepsilon _{mnop}\left( \overline{Q_{1_{L}}
}_{m}\rho _{n}\chi _{o}\eta _{p}^{\ast }\right) u_{a_{R}} +\frac{\lambda
_{ia}^{d}\text{ }}{\Lambda ^{2}}\varepsilon _{mnop}\left( \overline{Q_{i_{L}}
}_{m}\rho _{n}^{\ast }\chi _{o}^{\ast }\eta _{p}\right) d_{a_{R}}+h.c.\,,
\label{qeop}
\end{eqnarray}
where again, $i=2,\,3$ and $a=1,\,2,\,3$ label the family number (or generation). Considering these operators and the Yukawa lagrangian, Eq.~(\ref{yukawa}), 
the up-type quarks mass matrix, in the basis $(u_{1},$ $u_{2},$ $u_{3})$, is written as,
\begin{equation}
M_{u}=\frac{1}{\sqrt{2}}\left( 
\begin{array}{ccc}
\lambda _{11}^{u}\frac{v_{\chi }v_{\rho }v_{\eta }}{2\Lambda ^{2}} & \lambda
_{12}^{u}\frac{v_{\chi }v_{\rho }v_{\eta }}{2\Lambda ^{2}} & \lambda
_{13}^{u}\frac{v_{\chi }v_{\rho }v_{\eta }}{2\Lambda ^{2}} \\ 
\lambda _{21}^{u}v_{\rho } & \lambda _{22}^{u}v_{\rho } & \lambda
_{23}^{u}v_{\rho } \\ 
\lambda _{31}^{u}v_{\rho } & \lambda _{32}^{u}v_{\rho } & \lambda
_{33}^{u}v_{\rho }
\end{array}
\right) \,,
\label{upmatrix}
\end{equation}
while the down-type quarks mass matrix, in the basis ($d_{1},$ $d_{2},$ $d_{3})$, is 
\begin{equation}
M_{d}=\frac{1}{\sqrt{2}}\left( 
\begin{array}{ccc}
\lambda _{11}^{d}v_{\rho } & \lambda _{12}^{d}v_{\rho } & \lambda
_{13}^{d}v_{\rho } \\ 
\lambda _{21}^{d}\frac{v_{\chi }v_{\rho }v_{\eta }}{2\Lambda ^{2}} & \lambda
_{22}^{d}\frac{v_{\chi }v_{\rho }v_{\eta }}{2\Lambda ^{2}} & \lambda
_{23}^{d}\frac{v_{\chi }v_{\rho }v_{\eta }}{2\Lambda ^{2}} \\ 
\lambda _{31}^{d}\frac{v_{\chi }v_{\rho }v_{\eta }}{2\Lambda ^{2}} & \lambda
_{32}^{d}\frac{v_{\chi }v_{\rho }v_{\eta }}{2\Lambda ^{2}} & \lambda
_{33}^{d}\frac{v_{\chi }v_{\rho }v_{\eta }}{2\Lambda ^{2}}
\end{array}
\right) .
\label{dmatrix}
\end{equation}

It is remarkable that, by choosing $v_\eta \approx v_\chi \approx \Lambda$, all entries in these matrices are proportional to the SM symmetry breaking scale, $v_\rho = 246$~GeV, tuned by the dimensionless couplings. In other words, by appropriately tuning these couplings we can recover the SM results concerning quark masses and mixing without much effort. For example, if we take the above matrices as diagonal (just for the sake of illustration) we have:
\begin{eqnarray}
&&m_{u}\approx \lambda^{u}_{11} \frac{v_{\chi}v_{\eta}v_{\rho}}{2\sqrt{2}\Lambda^{2}},\;\;\;\; m_{c}\approx \lambda^{u}_{22}\frac{v_{\rho}}{\sqrt{2}}, 
\;\;\;\;\; m_{t}\approx \lambda^{u}_{33}\frac{v_{\rho}}{\sqrt{2}},\notag \\
&&m_{d}\approx \lambda^{d}_{11}\frac{v_{\rho}}{\sqrt{2}},\;\;\;\;\; m_{s}\approx \lambda^{d}_{22} \frac{v_{\chi}v_{\eta}v_{\rho}}{2\sqrt{2}\Lambda^{2}},
\;\;\;\; m_{b}\approx \lambda^{d}_{33} \frac{v_{\chi}v_{\eta}v_{\rho}}{2\sqrt{2}\Lambda^{2}},
\end{eqnarray}
where $m_{u}$, $m_{c}$, $m_{t}$, $m_{d}$, $m_{s}$ e $m_{b}$ are the masses for the up, charm, top, down, strange and bottom quarks, respectively. For $v_{\eta}=v_\chi=2.0$~TeV and $\Lambda=3.0$~TeV, and considering the known quark masses,
\begin{eqnarray}
&&m_{u}=0.5~MeV,\hspace{0.07cm}m_{d}=4.95~MeV,\hspace{0.07cm}m_{s}=1.26~MeV,\notag \\
&&m_{c}=1.26~GeV,\hspace{0.07cm}m_{b}=4.25~GeV,\hspace{0.07cm}m_{t}=173~GeV,
\end{eqnarray}
we obtain the following values for the respective Yukawa couplings,
$\lambda^{u}_{11}\approx 1.3\times 10^{-5}$, $\lambda^{u}_{22}\approx 7.2\times 10^{-3}$, $\lambda^{u}_{33}\approx 1.0$, $\lambda^{d}_{11}\approx 2.8\times 10^{-5}$,
$\lambda^{d}_{22}\approx 2.7\times 10^{-3}$, $\lambda^{d}_{33}\approx 0.11$.

In what concerns the exotic quarks, their masses are obtained purely from the Yukawa lagrangian, Eq.~(\ref{yukawa}), providing the mass for $U_1$ and $J_1$,
%
%
%
\begin{equation}
m_{U_{1}}=\lambda _{11}^{U}\frac{v_{\eta }}{\sqrt{2}}\text{ and }%
m_{J_{1}}=\lambda _{11}^{J}\frac{v_{\chi }}{\sqrt{2}}.
\label{U1J1mass}
\end{equation}
It also leads to the exotic quarks mass matrices in the basis $(
J_{2}^{ },$ $J_{3}^{ })$,
\begin{equation}
m_{J^{ }}=\frac{v_{\chi }}{\sqrt{2}}\left( 
\begin{array}{cc}
\lambda _{22}^{J^{ }} & \lambda _{23}^{J^{ }} \\ 
\lambda _{32}^{J^{ }} & \lambda _{33}^{J^{ }}
\end{array}
\right) \,,
\label{Jmatrix}
\end{equation}
and also in the basis ($D_{2},$ $D_{3})$,
\begin{equation}
m_{D}=\frac{v_{\eta }}{\sqrt{2}}\left( 
\begin{array}{cc}
\lambda _{22}^{D} & \lambda _{23}^{D} \\ 
\lambda _{32}^{D} & \lambda _{33}^{D}
\end{array}
\right)\,.
\label{Dmatrix}
\end{equation}
These matrices can be conveniently diagonalized, leaving no massless quarks in the spectrum. Instead, just for illustration, we can assume them diagonal and conservatively using, as reference, the lower values obtained by the CMS Collaboration~\cite{PDG} for a sequential fourth family ($m_D > 675$~GeV and $m_U > 625$~GeV, which we use for the exotically charged quarks as well),\footnote{Notice that the exotic quarks in this model are not sequential new ones, and would need a specific treatment in order to impose any realistic bound on their masses (see for example Ref.~\cite{AAE}.} imply that all Yukawa couplings for the exotic quarks have to be approximately bigger than 0.45.

As we saw earlier, charged leptons do not receive mass from Yukawa lagrangian, but there is an effective dimension-5 operator that does this job,
\begin{equation}
\frac{\kappa _{l}}{\Lambda }\left({L_{a_{L}}^{c}}\rho ^{\ast
}\right) \left( \chi ^{\dagger }L_{a_{L}}\right) +h.c.\,,
\label{dim5op}
\end{equation}
implying the following mass relation for charged leptons,
\begin{equation}
m_{l}=\frac{1}{2}k_{l}\frac{v_{\rho }v_{\chi }}{\Lambda }\,.
\label{chargedleptonmass}
\end{equation}
This term provides an SM-like mass term for charged leptons once we assume $v_\chi \approx \Lambda$, with $k_l$ playing the role of a Yukawa coupling. Considering the set of values for the mass scale parameter we have been using and plugging in the known masses for the leptons, we obtain the following values for the $k_l$ couplings:
$\kappa_{e}\approx 6.1\times 10^{-6}$, $\kappa_{\mu}\approx 1.3\times 10^{-3}$, 
$\kappa_{\tau}\approx 2.2\times 10^{-2}$,
which are comparable to those of SM.

As for the neutrinos, the lowest order effective operator that engender their masses is dimension-9,
\begin{eqnarray}
&&\frac{\kappa^a_{\nu }}{\Lambda ^{5}}\varepsilon _{ijkl}\varepsilon
_{mnop}\left({{\overline L_{ai_{L}}^{c}}}\rho _{j}\chi _{k}\eta _{l}\right)
\left( L_{am_{L}}\rho _{n}\chi _{o}\eta _{p}\right) 
+ h.c.\,, 
\label{numassop}
\end{eqnarray}
where we are supposing a diagonal basis for simplicity, yielding,
\begin{equation}
m^a_{\nu }=\frac{1}{4}k^a_{\nu }\frac{v_{\rho }^{2}v_{\chi }^{2}v_{\eta }^{2}}{
\Lambda ^{5}}\,.
\label{numass}
\end{equation}
With the scales as before, we see that neutrino masses are proportional to $v_\rho^2/\Lambda \approx 10$~GeV, which demands tiny values for $\kappa_\nu^a\simeq 5\times 10^{-10}$ for a neutrino of $0.5$~eV, not a great improvement if we compare with SM plus right handed neutrinos forming Dirac mass terms, but it accounts for sub-eV neutrino masses as well.

This procedure has shown to be effective in producing the desired fermion mass spectrum in the 341 model. Nevertheless, there is an unwanted side effect that has to be circumvented, it concerns the presence of effective operators that would trigger fast proton decay.
The lowest order effective dimension-8 operator that leads to proton
decay is,
\begin{equation}
\frac{C_{1}}{\Lambda ^{4}}\varepsilon _{ijkl}\left( \overline{Q_{1i_{L}}^{c}}
L_{1j_{L}}\chi _{k}\eta _{l}\right) \left( \overline{u_{1_{R}}^{c}}
d_{1_{R}}\right) +h.c.\,,
\label{protondecay}
\end{equation}
where an anti-symmetric color contraction is implicit. This operator contains the following term,
\begin{equation}
\frac{C_{1}}{\Lambda ^{4}}v_{\chi }v_{\eta }\overline{u_{_{L}}^{c}}e_{L}%
\overline{u_{_{R}}^{c}}d_{1_{R}}+h.c.\,,
\label{protondecayint}
\end{equation}
which yields the decay channel, $p^+\rightarrow \pi^0 + e^+$.
The simplest way to avoid proton decay is to impose a discrete $Z_{2}$ symmetry over the quark fields, such that,
\begin{equation}
Q_{k_{L}}\rightarrow -Q_{k_{L}},\text{ }q_{k_{R}}\rightarrow -q_{k_{R}}, \text{ }J_{k_{R}}\rightarrow -J_{k_{R}}, \text{ }U_{1_{R}}\rightarrow -U_{1_{R}}, \text{ }D_{j_{R}}\rightarrow -D_{j_{R}},
\end{equation}
that guarantees the proton stability at any order without compromising any of the results we have obtained.

\section{Conclusions}
\label{sec5}
In this work, we have built a 341 gauge model with a lesser scalar content than usual. The original version contains four scalar quartets and a decuplet, while we reduced this sector to only three scalar quartets. That was possible because we have computed the beta function for its abelian gauge coupling, showing that a Landau pole exists at a scale not greater than $\Lambda \approx 5$~TeV. This suggests that some underlying structure could emerge before that scale is reached, allowing us to use non renormalizable effective operators to provide the known mass spectrum for the fermions, which otherwise would need the presence of those extra scalars. 
It is amazing that such a low scale constrains this class of model at the current energies reachable at colliders such as LHC and/or ILC. Besides, a reduced spectrum makes it most attractive for phenomenological studies, competing with supersymmetry, which has not shown any sign of existence till now.

It has to be remarked that the existence of a cutoff scale as low as few TeV, may circumvent the long standing hierarchy problem. Actually, it would appear meaningless to address this problem once the theory enters into a strongly coupled regime at TeV scale, where we no longer expect that the SM could be sensitive to. 

Also, an interesting outcome of our scenario is that a 331 models with neutral fermions, when embedded in this 341 model, have to be faced as possessing this same Landau pole, a result that is not obvious in this class of 331 models, where the Landau pole is believed to appear much beyond the Planck scale~\cite{Dias}. This may have some impact on the phenomenology of such models like the impossibility of implementing neutrino mass under the assumption of existing an arbitrarily high cutoff energy scale for these models. On the other hand, knowing the advantages of these models concerning their DM candidates and Higgs phenomenology, we can guarantee the same outcome for our model, besides some new extra particles that can be investigated under the light of current experiments at LHC.

\section*{Acknowledgments}
This work was supported by Conselho Nacional de Desenvolvimento Cient\'{i}fico e Tecnol\'ogico - CNPq (A.G.D. grant 03094/2013-3, C.A.S.P. grant 306923/2013-0, P.S.R.S. grant 305390/2012-0),  A.G.D is also supported by the grant 2013/22079-8, S\~ao Paulo Research Foundation (FAPESP),
and Coordena\c c\~ao de Aperfei\c coamento de Pessoal de N\'{\i}vel Superior - CAPES (P.R.D.P. PhD program scholarship).

\appendix

\section{Scalar Mass Spectrum}
\indent
Here we derive the mass spectrum for the scalar fields. In order to do that we consider
the following shift in the neutral scalars by their VEVs,
\begin{eqnarray}
\eta _{1}^{0} &\rightarrow &\frac{1}{\sqrt{2}}\left( R_{\eta_1}+iI_{\eta_1}\right) , \\
\eta_2^{0} &\rightarrow &\frac{1}{\sqrt{2}}\left( v_{\eta }+R_{\eta_2
}+iI_{\eta_2}\right) , \\
\rho ^{0} &\rightarrow &\frac{1}{\sqrt{2}}\left( v_{\rho }+R_{\rho
}+iI_{\rho }\right) , \\
\chi ^{0} &\rightarrow &\frac{1}{\sqrt{2}}\left( v_{\chi }+R_{\chi
}+iI_{\chi }\right) .
\end{eqnarray}

We then obtain the conditions for the minimum of the scalar potential in Eq.(\ref{pot}),
\begin{eqnarray}
\mu _{1}^{2}+\lambda _{1}v_{\eta }^{2}+\frac{1}{2}\lambda _{4}v_{\rho }^{2}+
\frac{1}{2}\lambda _{5}v_{\chi }^{2}=0, \\
\mu _{2}^{2}+\lambda _{2}v_{\rho }^{2}+\frac{1}{2}\lambda _{4}v_{\eta }^{2}+
\frac{1}{2}\lambda _{6}v_{\chi }^{2}=0, \\
\mu _{3}^{2}+\lambda _{3}v_{\chi }^{2}+\frac{1}{2}\lambda _{5}v_{\eta }^{2}+
\frac{1}{2}\lambda _{6}v_{\rho }^{2}=0.
\label{constraints}
\end{eqnarray}
With these constraints we can build the CP-even neutral scalars mass matrix in the basis ($R_{\eta_2},\, R_{\rho },\,R_{\chi }$),
\begin{equation}
\frac{1}{2}\left( 
\begin{array}{ccc}
2\lambda _{1}v_{\eta }^{2} & \lambda _{4}v_{\eta }v_{\rho } & \lambda
_{5}v_{\eta }v_{\chi } \\ 
\lambda _{4}v_{\eta }v_{\rho } & 2\lambda _{2}v_{\rho }^{2} & \lambda
_{6}v_{\rho }v_{\chi } \\ 
\lambda _{5}v_{\eta }v_{\chi } & \lambda _{6}v_{\rho }v_{\chi } & 
2\lambda _{3}v_{\chi }^{2}
\end{array}
\right) .
\label{CPeven}
\end{equation}
According to the results in Section~\ref{sec2}, it is natural to assume $v_{\rho}^{2}/v_{\chi }^{2}\ll 1$, which we will use to simplify the computation of eigenvalues and eigenvectors for the CP-even mass matrix above. This diagonalization is performed by employing perturbation theory to the second order, leading to the following mass eigenvalues,
\begin{eqnarray*}
M_{H_{1}}^{2} &\approx &\lambda _{2}v_{\rho }^{2}+\frac{\lambda _{3}\lambda
_{4}^{2}+\lambda _{6}\left( \lambda _{1}\lambda _{6}-\lambda _{4}\lambda
_{5}\right) }{\lambda _{5}^{2}-4\lambda _{1}\lambda _{3}}v_{\rho }^{2}, \\
M_{H_{2}}^{2} &\approx &c_{1}v_{\chi}^2+c_{2}v_{\rho }^{2}\approx c_{1}v_{\chi}^2, \\
M_{H_{3}}^{2} &\approx &c_{3}v_{\chi }^{2}+c_{4}v_{\rho }^{2}\approx c_{3}v_{\chi }^{2},
\label{CP-even-mass}
\end{eqnarray*}
where
\begin{equation}
c_{1}=\frac{1}{2}\left( \lambda _{1}+\lambda _{3}-\sqrt{\left( \lambda _{1}-\lambda _{3}\right) ^{2}+\lambda _{5}^{2}}\right)
v_{\chi }^{2},
\end{equation}
\begin{equation}
c_{2}=\frac{\left[ \lambda _{4}\left( \lambda _{1}-\lambda _{3}-\sqrt{\left(
\lambda _{1}-\lambda _{3}\right) ^{2}+\lambda _{5}^{2}}\right) +\lambda
_{5}\lambda _{6}\right] ^{2}}{4c_{1} %
\left[\lambda _{5}^{2}-\left( \lambda _{1}-\lambda _{3}\right) \left(
\lambda _{3}-\lambda _{1}+\sqrt{\left( \lambda _{1}-\lambda _{3}\right)
^{2}+\lambda _{5}^{2}}\right) \right]},
\end{equation}
\begin{equation}
c_{3}=\frac{1}{2}\left( \lambda _{1}+\lambda _{3}+\sqrt{%
\left( \lambda _{1}-\lambda _{3}\right) ^{2}+\lambda _{5}^{2}}\right), 
\end{equation}
\begin{equation}
c_{4}=\frac{\left[\lambda _{4}\left( \lambda _{1}-\lambda _{3}+\sqrt{\left(
\lambda _{1}-\lambda _{3}\right) ^{2}+\lambda _{5}^{2}}\right) +\lambda
_{5}\lambda _{6}\right] ^{2}}{4c_{3} %
\left[ \lambda _{5}^{2}+\left( \lambda _{1}-\lambda _{3}\right) \left(
\lambda _{1}-\lambda _{3}+\sqrt{\left( \lambda _{1}-\lambda _{3}\right)
^{2}+\lambda _{5}^{2}}\right) \right] },
\end{equation}
while the respective eingenstates are obtained by considering first order perturbation theory only,
\begin{eqnarray}
H_{1} &\approx &R_\rho , \\
H_{2} &\approx &\frac{\lambda _{1}-\lambda _{3}-\sqrt{\left( \lambda
_{1}-\lambda _{3}\right) ^{2}+\lambda _{5}^{2}}}{\sqrt{\lambda
_{5}^{2}+\left( \lambda _{1}-\lambda _{3}-\sqrt{\left( \lambda _{1}-\lambda
_{3}\right) ^{2}+\lambda _{5}^{2}}\right) ^{2}}}R_{\eta_2}  \notag \\
&&+\frac{\lambda _{5}}{\sqrt{\lambda _{5}^{2}+\left( \lambda _{1}-\lambda
_{3}-\sqrt{\left( \lambda _{1}-\lambda _{3}\right) ^{2}+\lambda _{5}^{2}}%
\right) ^{2}}}R_{\chi }, \\
H_{3} &\approx &\frac{\lambda _{1}-\lambda _{3}+\sqrt{\left( \lambda
_{1}-\lambda _{3}\right) ^{2}+\lambda _{5}^{2}}}{\sqrt{\lambda
_{5}^{2}+\left( \lambda _{1}-\lambda _{3}+\sqrt{\left( \lambda _{1}-\lambda
_{3}\right) ^{2}+\lambda _{5}^{2}}\right) ^{2}}}R_{\eta_2}  \notag \\
&&+\frac{\lambda _{5}}{\sqrt{\lambda _{5}^{2}+\left( \lambda _{1}-\lambda
_{3}+\sqrt{\left( \lambda _{1}-\lambda _{3}\right) ^{2}+\lambda _{5}^{2}}%
\right) ^{2}}}R_{\chi }.  \notag
\label{CP-even-eigenstates}
\end{eqnarray}
Observe that the state $G_{1} = R_{\eta _1}$ is a Goldstone boson and $H_1$ is identified with the Higgs boson, once it is the only neutral scalar to get mass at the electroweak scale, besides coming from a doublet under $SU(2)_L$ symmetry similar to the SM one. The remaining scalars, $H_2$ and $H_3$, are heavier than the Higgs, with masses proportional to the 341 symmetry breaking scale.

The neutral CP-odd scalars, $I_{\eta_1},$ $I_{\eta_2},$ $I_{\rho }$ and $I_{\chi
}$, are all massless. Together with $G_1 = R_{\eta_1}$, they complete the set of five neutral Goldstone modes eaten by the five neutral massive gauge bosons.

As for the simply charged scalars, their mass matrices can be divided into two. The mass matrix in the basis ($\rho_{2}^{\pm}$, $\eta_{1}^{\pm }$), is given by,
\begin{equation}
\frac{1}{2}\left( 
\begin{array}{cc}
\lambda _{7}v_{\eta }^{2} & \lambda _{7}v_{\eta}v_{\rho}
\\ 
\lambda _{7}v_{\eta }v_{\rho } & \lambda _{8}v_{\rho}^{2} 
\end{array}
\right),
\label{matrixcharged1}
\end{equation}
which, upon diagonalization, leads to the mass eigenvalues,
\begin{eqnarray}
M_{G_{1}^{\pm }}^{2} &=&0, \\
M_{h_{1}^{\pm }}^{2} &=&\frac{\lambda _{7}}{2}\left( v_{\eta }^{2}+v_{\rho
}^{2}\right),
\label{charged1mass}
\end{eqnarray}
and the respective eigenvectors,
\begin{eqnarray}
G_{1}^{\pm } &=&-\frac{v_{\eta }}{\sqrt{v_{\eta }^{2}+v_{\rho }^{2}}}\eta
_{1}^{\pm }+\frac{v_{\chi }}{\sqrt{v_{\eta }^{2}+v_{\rho }^{2}}}\rho
_{2}^{\pm }, \\
h_{1}^{\pm } &=&\frac{v_{\rho }}{\sqrt{v_{\eta }^{2}+v_{\rho }^{2}}}\eta
_{1}^{\pm }+\frac{v_{\eta }}{\sqrt{v_{\eta }^{2}+v_{\rho }^{2}}}\rho
_{2}^{\pm }\,.
\label{charged1states}
\end{eqnarray}
The second mass matrix, in the basis ($\eta_{2}^{\pm}$, $\chi_{2}^{\pm }$), is
\begin{equation}
\frac{1}{2}\left(
\begin{array}{cc}
\lambda _{8}v_{\chi}^{2} & \lambda _{8}v_{\eta }v_{\chi }
\\ 
\lambda _{8}v_{\eta }v_{\chi } & \lambda _{8}v_{\eta }^{2}%
\end{array}
\right)\,,
\label{matrixcharged2}
\end{equation}
yielding the mass eigenvalues,
\begin{eqnarray}
M_{G_{2}^{\pm }}^{2} &=&0, \\
M_{h_{2}^{\pm }}^{2} &=&\frac{\lambda _{8}}{2}\left(v_{\eta }^{2}+v_{\chi
}^{2}\right)\,,
\label{charged2mass}
\end{eqnarray}
and respective eigenvectors,
\begin{eqnarray}
G_{2}^{\pm } &=&-\frac{v_{\eta }}{\sqrt{v_{\eta }^{2}+v_{\chi }^{2}}}\eta
_{2}^{\pm }+\frac{v_{\chi }}{\sqrt{v_{\eta }^{2}+v_{\chi }^{2}}}\chi
_{2}^{\pm }, \\
h_{2}^{\pm } &=&\frac{v_{\chi }}{\sqrt{v_{\eta }^{2}+v_{\chi }^{2}}}\eta
_{2}^{\pm }+\frac{v_{\eta }}{\sqrt{v_{\eta }^{2}+v_{\chi }^{2}}}\chi
_{2}^{\pm }.
\label{charged2states}
\end{eqnarray}
In the above results $G_{1}^{\pm }$ and $
G_{2}^{\pm }$, together with  $\rho_{1}^{\pm}$ and $\chi_{1}^{\pm}$ (which are already massless eigenstates), are all massless and represent the Goldstone bosons eaten by the eight simply charged gauge bosons,
while the scalar fields, $h_{1}^{\pm }$ and $h_{2}^{\pm }$ are massive and remain in the physical spectrum.

Concerning the doubly charged scalars, we have the mass matrix in the basis $\left( \rho
^{\pm \pm },\chi ^{\pm \pm }\right) $,
\begin{equation}
\frac{\lambda _{9}}{2}\left( 
\begin{array}{cc}
v_{\chi }^{2} & v_{\rho }v_{\chi } \\ 
v_{\rho }v_{\chi } & v_{\rho }^{2}
\end{array}
\right)\,,
\label{matrixdcharged1}
\end{equation}
that leads to a null eigenvalue, $M_{G^{\pm\pm}}=0$, and
\begin{equation}
M_{h^{\pm \pm }}^{2}=\frac{\lambda _{9}}{2}\left( v_{\rho }^{2}+v_{\chi
}^{2}\right)\,,
\label{massdcharged}
\end{equation}
whose eigenvectors are,
\begin{eqnarray}
G^{\pm \pm } &=&-\frac{v_{\rho }}{\sqrt{v_{\rho }^{2}+v_{\chi }^{2}}}\rho
^{\pm \pm }+\frac{v_{\chi }}{\sqrt{v_{\rho }^{2}+v_{\chi }^{2}}}\chi ^{\pm
\pm }, \\
h^{\pm \pm } &=&\frac{v_{\chi }}{\sqrt{v_{\rho }^{2}+v_{\chi }^{2}}}\rho
^{\pm \pm }+\frac{v_{\rho }}{\sqrt{v_{\rho }^{2}+v_{\chi }^{2}}}\chi ^{\pm
\pm }\,.
\end{eqnarray}
The scalar $G^{\pm \pm }$ is obviously a Goldstone boson eaten by the doubly charged gauge boson, while $h^{\pm \pm }$ remains in the spectrum, being a characteristic signature in this class of models. Below we present the numerical values of couplings chosen to give a Higgs mass of 126~GeV as well as the masses of all scalars when $v_\eta = v\chi =2$~TeV.
\begin{eqnarray}
\lambda _{1}=0.25,\ \lambda _{2}=0.28,\ \lambda _{3}=0.18,\ \lambda _{4}=0.1 \\
\lambda _{5}=0.15,\ \lambda _{6}=0.1,\ \lambda _{7}=0.2,\ \lambda _{8}=0.21,\ \lambda _{9}=0.23.
\end{eqnarray}
which yield the CP-even scalar masses,
\begin{eqnarray*}
M_{H_{1}} &\approx &126\text{ }GeV, \\
M_{H_{2}} &\approx &727\text{ }GeV, \\
M_{H_{3}} &\approx &1092\text{ }GeV,
\end{eqnarray*}
while the simply charged scalars acquired the following masses.
\begin{eqnarray}
M_{h_{1}^{\pm }} &\approx &632\text{ }GeV, \\
M_{h_{2}^{\pm }} &\approx &916\text{ }GeV.
\end{eqnarray}
Finally, the doubly charged scalar gets a mass,
\begin{equation}
M_{h^{\pm \pm }}\approx 959\text{ }GeV.
\end{equation}
We have to keep in mind that the above particles have non-standard couplings to fermions and it is not straightforward to
compare them with the existing bounds from colliders, although they are heavy enough to be safe and tested at the next LHC run. A more careful and thorough analysis should be made elsewhere.

\section{Symmetry breakdown in matter and scalar Sector}
\indent

The effect of 341 to 331 symmetry breaking driven by the VEV 
$v_\chi$, causes fermionic quartets to decompose into triplets plus
singlets. 
The 341 model presented here contains as a subgroup the 331 model with neutral fermions in the third component of the lepton triplets. There are two possibilities for this extra neutral fermion (one for each family), it can be the partner right-handed neutrino of the left-handed neutrino in the triplet, or it can be a new neutral fermion not related to the ordinary neutrino. In order to see that, let us decompose the content of the 341 model into multiplets of 331 in the following way:
\begin{itemize}
\item{Leptons}
{\small \begin{equation}
L_{a_{L}}=\left( 
\begin{array}{c}
\nu _{a} \\ 
e_{a} \\ 
\nu _{a}^{c} \\ 
e_{a}^{c}
\end{array}
\right) _{L}\longrightarrow L_{a_{L}}^{\prime }=\left( 
\begin{array}{c}
\nu _{a} \\ 
e_{a} \\ 
\nu _{a}^{c}
\end{array}
\right) _{L}\sim \left( \mathbf{1},\mathbf{3},-1/3\right)
\oplus e_{aL}^{c}\sim \left( \mathbf{1},\mathbf{1},+1\right)\,,
\label{declep}
\end{equation}}
where, from now on, the transformation properties under parentheses refer to the 331 symmetry.
\item{First quark generation} 
{\small \begin{equation}
Q_{1L}=\left( 
\begin{array}{c}
u_{1} \\ 
d_{1} \\ 
U_{1} \\ 
J_{1}
\end{array}
\right) _{L}\longrightarrow Q_{1_{L}}^{\prime }=\left( 
\begin{array}{c}
u_{1} \\ 
d_{1} \\ 
U_{1}
\end{array}
\right) _{L}\sim \left( \mathbf{3},\mathbf{3},+1/3\right)
\oplus J_{1_{R}}\sim \left( \mathbf{3},\mathbf{1},+5/3\right) \,.
\label{decquark1}
\end{equation}}
\item{Second and third quark generations}
{\small \begin{equation}
Q_{iL}=\left( 
\begin{array}{c}
d_{i} \\ 
u_{i} \\ 
D_{i} \\ 
J_{i}^{ }
\end{array}
\right) _{L}\longrightarrow Q_{i_{L}}^{\prime }=\left( 
\begin{array}{c}
d_{i} \\ 
u_{i} \\ 
D_{i}
\end{array}
\right) _{L}\sim \left( \mathbf{3},\mathbf{3}^{\ast },0\right) \oplus J_{i_{R}}^{ }\sim \left( \mathbf{3},\mathbf{1}
,-4/3\right)\,.
\label{decquark2}
\end{equation}}
\item{Scalars}
{\small \begin{eqnarray*}
\chi &=&\left( 
\begin{array}{c}
\chi _{1}^{-} \\ 
\chi ^{--} \\ 
\chi _{2}^{-} \\ 
\chi ^{0}
\end{array}
\right) \longrightarrow \text{ }\chi ^{\prime }=\left( 
\begin{array}{c}
\chi _{1}^{-} \\ 
\chi ^{--} \\ 
\chi _{2}^{-}
\end{array}
\right) \sim \left( \mathbf{1},\mathbf{3},-4/3\right) \oplus \chi
^{0}\sim \left( \mathbf{1},\mathbf{1},0\right) , \\
\eta &=&\left( 
\begin{array}{c}
\eta _{1}^{0} \\ 
\eta _{1}^{-} \\ 
\eta _{2}^{0} \\ 
\eta _{2}^{+}
\end{array}
\right) \longrightarrow \text{ }\eta ^{\prime }=\left( 
\begin{array}{c}
\eta _{1}^{0} \\ 
\eta _{1}^{-} \\ 
\eta _{2}^{0}
\end{array}
\right) \sim \left( \mathbf{1},\mathbf{3},-1/3\right) \oplus \eta
_{2}^{+}\sim \left( \mathbf{1},\mathbf{1},+1\right) , \\
\rho &=&\left( 
\begin{array}{c}
\rho _{1}^{+} \\ 
\rho ^{0} \\ 
\rho _{2}^{+} \\ 
\rho ^{++}
\end{array}
\right) \rightarrow \rho ^{\prime }=\left( 
\begin{array}{c}
\rho _{1}^{+} \\ 
\rho ^{0} \\ 
\rho _{2}^{+}
\end{array}
\right) \sim \left( \mathbf{1},\mathbf{3},+2/3\right) \oplus \rho
^{++}\sim \left( \mathbf{1},\mathbf{1},+2\right) \,.
\label{decscalars}
\end{eqnarray*}}
\end{itemize}

Next, we present the breaking of 331 model promoted by $v_\eta$ that leads to the 321 SM content plus extra singlets.
For the leptons we have,
\begin{equation}
L_{a_{L}}^{\prime }\longrightarrow l_{a_{L}}=\left( 
\begin{array}{c}
\nu _{a} \\ 
e_{a}
\end{array}
\right) _{L}\sim \left( \mathbf{1},\mathbf{2},Y_{l_{a}}\right) \oplus \nu
_{aL}^{c}\sim \left( \mathbf{1},\mathbf{1},Y_{\nu _{a}}\right) ,
\end{equation}
where $Y_{l_{a}}=-1$ and $Y_{\nu _{a}}=0.$

For the first quark generation we have,
\begin{equation}
Q_{1L}^{\prime }\rightarrow q_{1L}=\left( 
\begin{array}{c}
u_{1} \\ 
d_{1}
\end{array}
\right) _{L}\sim \left( \mathbf{3},\mathbf{2},Y_{q_{1}}\right)
\oplus U_{1_{R}}\sim \left( \mathbf{3},\mathbf{1},Y_{U_{1}}\right)
\end{equation}
where $Y_{q_{1}}=\frac{1}{3}$ and $Y_{U_{1}}=\frac{4}{3}$.

For the second and third quark generations, we have 
\begin{equation}
Q_{iL}^{\prime }\rightarrow \widetilde{q}_{iL}=\left( 
\begin{array}{c}
d_{i} \\ 
u_{i}
\end{array}
\right) _{L}\sim \left( \mathbf{3},\mathbf{2},Y_{q_{i}}\right)
\oplus D_{i_{R}}\sim \left( \mathbf{3},\mathbf{1},Y_{D_{i}}\right) ,
\end{equation}
where $Y_{q_{i}}=\frac{1}{3}$ and $Y_{D_{i}}=-\frac{2}{3}$.

The scalar triplets decompose into one doublet and one singlet each, 
\begin{eqnarray}
\text{ }\chi ^{\prime } &\rightarrow &\chi ^{\prime \prime }=\left( 
\begin{array}{c}
\chi _{1}^{-} \\ 
\chi ^{--}
\end{array}
\right) \sim \left( \mathbf{1},\mathbf{2},Y_{\chi ^{\prime \prime }}\right)
\oplus \chi _{2}^{-}\sim \left( \mathbf{1},\mathbf{1},Y_{\chi _{2}^{-}}\right) , \\
\eta ^{\prime } &\rightarrow &\eta ^{\prime \prime }=\left( 
\begin{array}{c}
\eta _{1}^{0} \\ 
\eta _{1}^{-}
\end{array}
\right) \text{ }\sim \left( \mathbf{1},\mathbf{2},Y_{\eta ^{\prime \prime
}}\right) \oplus \eta _{2}^{0}\sim \left( \mathbf{1},\mathbf{1},Y_{\eta
_{2}^{0}}\right) , \\
\rho ^{\prime } &\rightarrow &\rho ^{\prime \prime }=\left( 
\begin{array}{c}
\rho _{1}^{+} \\ 
\rho ^{0}
\end{array}
\right) \sim \left( \mathbf{1},\mathbf{2},Y_{\rho ^{\prime \prime }}\right)
\oplus \rho _{2}^{+}\sim \left( \mathbf{1},\mathbf{1},Y_{\rho _{2}^{+}}\right)\,,
\end{eqnarray}
where $Y_{\chi ^{\prime \prime }}=-3,$ $Y_{\eta ^{\prime \prime }}=-1,$ $%
Y_{\rho ^{\prime \prime }}=1,$ $Y_{\chi _{2}^{-}}=-2,$ $Y_{\eta _{2}^{0}}=0$
and $Y_{\rho _{2}^{+}}=2$. The scalar doublet $\rho ^{\prime \prime }$ plays
the role of the Higgs doublet in the SM. So, we recover all the effective SM doublets and singlets
with their respective quantum numbers, plus right-handed neutrinos, extra quarks and scalars.

\section{Gauge bosons mass spectrum}

\indent

The gauge bosons obtain their masses from the Lagrangian, 
\begin{equation}
\mathcal{L=}\left( \mathcal{D}_{\mu }\chi \right) ^{\dagger }\left( \mathcal{%
D}^{\mu }\chi \right) +\left( \mathcal{D}_{\mu }\eta \right) ^{\dagger
}\left( \mathcal{D}^{\mu }\eta \right) \mathcal{+}\left( \mathcal{D}_{\mu
}\rho \right) ^{\dagger }\left( \mathcal{D}^{\mu }\rho \right)\,,
\label{GBkinectic}
\end{equation}
which, after spontaneous symmetry breakdown, leads to the
charged (and non-Hermitian) gauge boson masses,
\begin{eqnarray}
W^{\pm } &=&\frac{W_{\mu }^{1}\mp iW_{\mu }^{2}}{\sqrt{2}}\rightarrow
M_{W^{\pm }}^{2}\mathcal{=}\frac{1}{4}g_{L}^{2}v_{\rho }^{2},\text{ } \approx (80~{\mbox GeV})^2\\
K^{0},K^{\prime 0} &=&\frac{W_{\mu }^{4}\pm iW_{\mu }^{5}}{\sqrt{2}}
\rightarrow M_{K^{0},K^{\prime 0}}^{2}\mathcal{=}\frac{1}{4}g_{L}^{2}v_{\eta
}^{2},\text{ } \approx (650~{\mbox GeV})^2\\
K_{1}^{\pm } &=&\frac{W_{\mu }^{6}\pm iW_{\mu }^{7}}{\sqrt{2}}\rightarrow
M_{K_{1}^{\pm }}^{2}\mathcal{=}\frac{1}{4}g_{L}^{2}\left( v_{\eta
}^{2}+v_{\rho }^{2}\right) \approx (655~{\mbox GeV})^2
\\
X^{\pm } &=&\frac{W_{\mu }^{9}\pm iW_{\mu }^{10}}{\sqrt{2}}\rightarrow
M_{X^{\pm }}^{2}\mathcal{=}\frac{1}{4}g_{L}^{2}v_{\chi }^{2} \approx (650~{\mbox GeV})^2
\\
V^{\pm \pm } &=&\frac{W_{\mu }^{11}\pm iW_{\mu }^{12}}{\sqrt{2}}\rightarrow
M_{V^{\pm \pm }}^{2}\mathcal{=}\frac{1}{4}g_{L}^{2}\left( v_{\rho
}^{2}+v_{\chi }^{2}\right) \approx (655~{\mbox GeV})^2
\\
Y^{\pm } &=&\frac{W_{\mu }^{13}\pm iW_{\mu }^{14}}{\sqrt{2}}\rightarrow
M_{Y^{\pm }}^{2}\mathcal{=}\frac{1}{4}g_{L}^{2}\left( v_{\eta }^{2}+v_{\chi
}^{2}\right)  \approx (920~{\mbox GeV})^2
\,.
\label{chargedGBmass}
\end{eqnarray}

It also leads to the mass matrix for the neutral gauge bosons, in the basis ($W^{3},$ $W^{8},$ $W^{15},$ $W^{X}$), 
\begin{equation}
\frac{g_{L}^{2}}{4}\left( 
\begin{array}{cccc}
v_{\rho }^{2} & -\frac{1}{\sqrt{3}}v_{\rho }^{2} & -\frac{1}{\sqrt{6}}
v_{\rho }^{2} & -2tv_{\rho }^{2} \\ 
-\frac{1}{\sqrt{3}}v_{\rho }^{2} & \frac{1}{3}\left( v_{\rho }^{2}+4v_{\eta
}^{2}\right) & \frac{1}{3\sqrt{2}}\left( v_{\rho }^{2}-2v_{\eta }^{2}\right)
& \frac{2t}{\sqrt{3}}v_{\rho }^{2} \\ 
-\frac{1}{\sqrt{6}}v_{\rho }^{2} & \frac{1}{3\sqrt{2}}\left( v_{\rho
}^{2}-2v_{\eta }^{2}\right) & \frac{1}{6}\left( v_{\eta }^{2}+v_{\rho
}^{2}+9v_{\chi }^{2}\right) & \frac{2t}{\sqrt{6}}\left( v_{\rho
}^{2}+3v_{\chi }^{2}\right) \\ 
-2tv_{\rho }^{2} & \frac{2t}{\sqrt{3}}v_{\rho }^{2} & \frac{2t}{\sqrt{6}}
\left( v_{\rho }^{2}+3v_{\chi }^{2}\right) & 4t^{2}\left( v_{\rho
}^{2}+v_{\chi }^{2}\right)
\end{array}
\right)\,,
\label{massmatrixNGB}
\end{equation}
where $t\equiv \frac{g_{X}}{g_{L}}$. This mass
matrix has determinant equal to zero, which guarantees the existence of a massless gauge boson, that we can associate with the photon. Diagonalizing the matrix using the simplifying (and reasonable)
assumption $v_{\chi }\approx v_{\eta }\gg v_{\rho }$, we obtain
the neutral gauge boson masses, as given below: 
\begin{eqnarray}
M_{A}^{2} &=&0,\text{ } \\
M_{Z}^{2} &=&\frac{g_L^{2}v_{\rho }^{2}}{4c_{W}^{2}} = \frac{M_{W^\pm}^2}{c_W^2}\text{ } \approx (91~{\mbox GeV})^2 \\
M_{Z^{\prime }}^{2} &\approx &\frac{g_L^{2}c_W^2v_{\eta }^{2}}{h_{W}} \approx (790~{\mbox GeV})^2, \\
M_{Z^{^{\prime \prime }}}^{2} &\approx &\frac{g_L^{2}v_{\eta }^{2}\left[ \left(
1-4s_{W}^{2}\right) ^{2}+h_{W}^{2}\right] }{8h_{W}\left(
1-4s_{W}^{2}\right)} \approx (2.2~{\mbox TeV})^2\,,
\label{neutralGBmass}
\end{eqnarray}
with the respective eigenstates,
\begin{eqnarray}
A^{\mu } &=& S_{W}W_{3}^{\mu }+C_{W}\left[ \frac{T_{W}}{\sqrt{3}}\left(
-W_{8}^{\mu }-2\sqrt{2}W_{15}^{\mu }\right) +\sqrt{1-3T_{W}^{2}}W_{X}^{\mu }%
\right] , \\
Z^{\mu } &=& C_{W}W_{3}^{\mu }-S_{W}\left[ \frac{T_{W}}{\sqrt{3}}\left(
-W_{8}^{\mu }-2\sqrt{2}W_{15}^{\mu }\right) +\sqrt{1-3T_{W}^{2}}W_{X}^{\mu }%
\right] , \\
Z^{\prime \mu } &=&\frac{\sqrt{3}}{3}\frac{\sqrt{1-3T_{W}^{2}}}{\sqrt{%
1-4S_{W}^{2}}}\left( \sqrt{h_{W}}W_{8}^{\mu }-2\sqrt{2}\frac{S_{W}^{2}}{%
\sqrt{h_{W}}}W_{15}^{\mu }\right) +\frac{S_{W}\sqrt{1-3T_{W}^{2}}}{\sqrt{%
h_{W}}}W_{X}^{\mu }, \nonumber \\ \\
Z^{\prime \prime \mu } &=&\frac{\sqrt{3}\sqrt{1-4S_{W}^{2}}}{\sqrt{h_{W}}}%
W_{15}^{\mu }+\frac{2\sqrt{2}S_{W}}{\sqrt{h_{W}}}W_{X}^{\mu },
\end{eqnarray}
In the above equations we have used the sine of the electroweak mixing angle written as $s_{W}=t/\sqrt{1+4t^{2}}$, and defined $\cos{\theta_W} \equiv c_W$, $\tan{\theta_W} \equiv t_W$ and $h_W\equiv 3-4s_W^2$. 
 From these results we see that $M_{Z^{\prime \prime}}>M_{Z^{\prime }} >M_{Z}$, recovering the $Z^{\prime }$ of the $331_{RHN}$~\cite{341vicente,FP}
model, and the $Z$ boson reproduces the well established massive neutral gauge boson of SM. 
We then have an extra neutral gauge boson, $Z^{\prime \prime }$, which is heavier than $Z^\prime$, whose phenomenology may easily be probed at LHC and/or the next collider generation.

\end{document}